\documentclass[aps,prb,preprint,showpacs]{revtex4}
\usepackage{graphicx}% Include figure files
\usepackage{dcolumn}% Align table columns on decimal point
\usepackage{bm}% bold math

\newcommand{\bcmo}{Bi$_{1-x}$Ca$_{x}$MnO$_{3}$ }
\newcommand{\opps}{pO$_{2}$ }
\newcommand{\oppsb}{pO$_{2}$}
\newcommand{\td}{t$_{d}$}
\newcommand{\tds}{t$_{d}$ }
\newcommand{\tw}{t$_{w}$}

\begin{document}

\title{Studies of structural, magnetic, electrical  and photoconducting  properties
of \bcmo epitaxial thin films. }

\author{S. Chaudhuri}
\author{R. C. Budhani}
\email{rcb@iitk.ac.in} \affiliation{Department of Physics \\
Indian Institute of Technology Kanpur \\ Kanpur - 208016, India}

\date{\today}

\date{\today}% It is always \today, today,

\begin{abstract}
The dynamics of the charge ordered (CO) state under
non-equilibrium conditions created by strong dc-electric field
($\leq$ $10^{6}$ V/cm) and photo-illumination with short
($\approx$ 6 ns) laser pulses is investigated in \bcmo (x $>$ 0.5)
epitaxial films. A pulsed laser deposition method was used to
synthesize films on (100) LaAlO$_{3}$ (LAO) and (100) SrTiO$_{3}$
(STO) substrates. The crystallographic structure, temperature
dependence of electrical resistivity and magnetization of the
samples of different composition prepared under different oxygen
partial pressure (pO$_{2}$) and deposition temperature (T$_{D}$)
are studied. For the x = 0.6 sample grown on LAO, a clear
signature of charge ordering at $\approx$ 275 K is seen in the
magnetization and at $\approx$ 260 K in the resistivity data. The
same sample grown on STO revealed a complex behavior, which
entails charge ordering at $\approx$ 300 K, a N$\acute{e}$el order
at $\approx$ 150 K and finally a weak ferromagnetic phase below 50
K. A strong correlation between  charge ordering temperature
(T$_{CO}$) and the c-axis lattice parameter (\textit{c}) of the
type ( dT$_{CO}$/dc $\approx$ -350 K/{\AA} ) imerges from
measurements on films deposited under different growth conditions.
Since the out of plane lattice parameter (\textit{c}) increases
with in plane compressive strain, this effect directly show a
compressive strain induced suppression of the T$_{CO}$. The
current (I)- voltage (V) characteristics of the samples at T $<$
T$_{CO}$ show hysteresis due to a compound effect of Joule heating
and collapse of the CO state. Transient changes in conductivity of
lifetime ranging from nano to microseconds are seen at T $<$
T$_{CO}$ on illumination with pulsed  UV (355 nm) radiation. These
observations are explained on the basis of the topological and
electronic changes in the charge ordered phase.
\end{abstract}
 \maketitle

\section{Introduction}
The manganite Bi$_{1-x}$Ca$_{x}$MnO$_{3}$ (BCMO) is a fairly well
studied system in the bulk ceramic and single crystal forms.
\cite{ref1,ref2,ref3,ref4,ref5,ref6,ref7} BCMO is known to exhibit
Charge Ordering (CO) over a much broader range of hole doping than
its isostructural counterpart, La$_{1-x}$Ca$_{x}$MnO$_{3}$. But
the key property that makes BCMO stand out, atleast from the
technological view point, is its behavior upon photoillumination.
Manganites are known to exhibit the coexistence of a charge
ordered insulating (COI) and charge delocalized ferromagnetic
metallic (CDFM) phases under the non-equilibrium conditions
created by photoillumination, whereby regions exposed to the
photon field undergo a transition from the initially insulating to
a metallic state leaving the unexposed regions unaffected.
\cite{ref8,ref9,ref10,ref11,ref12,ref13,ref14,ref15,ref16,ref17,ref18}
Owing to the difference in their electrical conductivity, the
exposed and the unexposed regions have different refractive
indices. This feature, along with the localized nature of the
photoinduced insulator-to-metal transition (IMT), can be exploited
for the creation of photonic bandgap structures using holographic
techniques. \cite{ref6} But in most of the manganites this effect
is transient as the pressure on the conducting phase from the
surrounding insulating matrix forces it back to its insulating
state, and makes the practical realization of manganites as
photonic bandgap crystals a challenging issue. However, unlike the
other charge ordered manganites, BCMO has been found to exhibit a
persistent change in properties on photoillumination$.\cite{ref6}$
Owing to this extraordinary property, BCMO is a promising
candidate for photonic crystals. BCMO is well studied in its bulk
form, Woo \textit{et al.}\cite{ref3} have performed resistivity,
magnetization, X-ray absorption as well as X-ray diffraction
measurements to correlate the structural, magnetic and transport
properties of BCMO. They reported the charge ordering temperature
T$_{CO}$ for x = 0.4, 0.6, and 0.8 to be 315 K, 330 K, and 190 K
and the N$\acute{e}$el temperature T$_{N}$ for x = 0.4, and 0.6,
to be 150 K, and 125 K respectively, and point out the importance
of Mn-O bond distortion in stabilizing the CO phase. Bokov
\textit{et al.}\cite{ref1} have given a detailed account of the
structural properties along with electrical transport and magnetic
properties for 0.2 $\leq$ x $\leq$ 1. Podzorov $\textit{et
al.}$$\cite{ref5}$ found that the transformation to the charge
ordered state to be martensitic in character. This non-diffusive
transformation by its very nature, leads to long-range deformation
of the crystal lattice and results in accommodation strain at the
charge ordered domain boundaries. Since, the martensitic strain is
different from the substrate induced lattice deformation, a
compound effect of these two strains in thin film samples can
either enhance or suppress the properties seen in bulk single
crystals. These issues make studies on epitaxial thin films of
BCMO important. So far only a limited study of photoconductivity
in thin films of Bi$_{1-x}$Ca$_{x}$MnO$_{3}$ on illumination with
Ar-ion CW laser of mixed wavelength has been reported.
\cite{ref24}

This paper presents a detailed investigation of the growth
conditions, structure, electrical transport, magnetic ordering and
transient photoconductivity on pulsed laser illumination on three
different compositions of Bi$_{1-x}$Ca$_{x}$MnO$_{3}$ films in the
overdoped regime. Studies on thin films are desirable as any
conceivable practical application of photoinduced effects and
related metal-insulator transition require the material in a thin
film form. Furthermore, an epitaxial film supported on a substrate
is likely to have fundamentally different properties due to
quantum size effects and substrate related changes in stress and
growth morphology. \cite{ref19,ref26,ref32} With this in mind, we
have prepared epitaxial thin films of \bcmo with x = 0.6, 0.67 and
0.76 on STO (100) and LAO (100). The deposition temperature and
the oxygen partial pressure were optimized to obtain films of good
crystalline quality and the variation of the lattice parameters as
a function of these growth parameters were studied by X-ray
diffraction. Electrical resistivity and magnetization measurements
on the x = 0.6 sample reveal clear signatures of charge ordering
at $\approx$ 260 K and $\approx$ 275 K respectively.  The electric
field dependence of charge transport in the CO - state is highly
non-linear and shows history effects, whose origin is both thermal
and electronic. The photoresponse measurements on x = 0.6 sample
exhibit a transient conducting state which can be attributed to
the melting of the charge ordered state followed by freezing to
the original insulating state over time scale of several
microseconds. The measurement of current transport as well as
photoeffects reveal a robust CO - state in BCMO.

\section{Experimental}
Thin films of thickness $\approx$ 2000  {\AA} of \bcmo with x =
0.6, 0.67 and 0.76 were grown at the rate of $\approx$ 1 {\AA}/sec
on (100) oriented single crystals of LaAlO$_{3}$ (rhombohedral
with a = 3.788  {\AA}) and  SrTiO$_{3}$ (cubic with a = 3.905
{\AA}) using pulsed laser deposition (PLD) technique. A KrF
excimer laser operated at 10 Hz with an areal energy density of 2
J/cm$^{2}$/pulse on the surface of a stoichiometric sintered
target of BCMO was used for ablation. The film growth conditions
were optimized by varying the deposition temperature (T$_{D}$) and
oxygen partial pressure (\oppsb) to realize a sharp step in
electrical resistivity at T$_{CO}$. The films with the most
prominent signature of charge ordering were obtained for T$_{D}$
and \opps of $\approx$ 800 $^{O}$C and 350 mTorr respectively by
ablation of the target of x = 0.6. The crystallographic structure
and the surface roughness of the films were probed with X-ray
$\Theta$$\--$2$\Theta$ diffraction and atomic force microscopy
(AFM) techniques respectively. The AFM measurements revealed a
smooth surface of roughness $<$ 25 {\AA}. All magnetization
measurements were carried out using a SQUID magnetometer [Quantum
Design MPMS XL5]. Measurements of electrical resistivity,
current-voltage characteristics (I-V) and photoinduced changes in
resistivity were carried out on 1 mm wide strips of the film on
which silver pads were deposited through shadow mask for
connecting current and voltage leads. We have made bridges of
length $\geq$ 1 $\mu$m by using silica fibers as masks. The
standard four probe technique was used to measure the electrical
resistivity. Short pulses ( $\approx$ 6 ns ) of the third harmonic
frequency ($\lambda$ = 355 nm) of an Nd:YAG laser were used for
photoexcitation experiments. Changes in the voltage drop across
the sample on photoillumination were captured using a fast digital
oscilloscope (Tektronix TDS-380) with a time resolution of 0.5 ns.

\section{RESULTS AND DISCUSSION}
\subsection{Structural characterization}
In Fig. 1 we show the X-ray diffraction profiles of the samples
with x = 0.6 grown at different \opps but at the same T$_{D}$ of
750 $^{O}$C. The diffractograms are dominated by the strong (002)
reflection of LaAlO$_{3}$ next to which a small shoulder
corresponding to (002) reflection of the BCMO appears. This
feature has been marked by an arrow in the figure. The inset of
Fig. 1 shows the variation of the out-of-plane lattice parameter
(\textit{c}) as the \opps is changed. For the sample deposited at
\opps of 350 mTorr, the \textit{c} is $\approx$ 3.81  {\AA}, in
close agreement with the bulk value $\approx$ 3.82  {\AA}
calculated by Bokov \textit{et al}. \cite{ref1} X-ray diffraction
measurements on samples deposited on LAO at \opps = 350 mTorr and
different T$_{D}$ reveal that the crystalline quality of the thin
films degrades on lowering the T$_{D}$ below $\approx$ 700
$^{O}$C. The X-ray diffraction patterns for samples of different
composition grown at T$_{D}$ = 750 $^{O}$C and \opps = 350 mTorr
are shown in Fig 2. The c-axis lattice parameter is found to
decrease with increasing calcium content, as observed in bulk
samples. \cite{ref1} This compositional variation of the lattice
parameter is due to the fact that the end members BiMnO$_{3}$ and
CaMnO$_{3}$  have monoclinic (with \textit{a} = \textit{c} = 3.932
{\AA}, \textit{b} = 3.989 {\AA}) and cubic (\textit{a} = 3.725
{\AA}) structures respectively. For the sake of comparison, in
Fig. 3 we show the diffraction pattern of the sample grown on STO
at \opps of 350 mTorr and T$_{D}$ of 750 $^{O}$C. The out-of-plane
lattice parameter \textit{c} is 3.75 {\AA} in this case, which is
less by $\approx$ 2 $\%$ as compared to the film deposited on LAO
under similar conditions. Clearly the growth of BCMO films on STO
puts a tensile strain on the a-b plane of the BCMO, and in order
to maintain the unit cell volume the c-axis undergoes a
contraction.

\subsection{Magnetic ordering}
Figure 4 shows the temperature dependence of magnetization (M)
measured at 0.5 tesla in-plane field in field-cooled (FC) and zero
field-cooled (ZFC) modes for the x = 0.6 sample grown at 750
$^{O}$C and \opps of 350 mTorr. The maxima in the magnetization at
$\approx$  275 K is similar to the observations of Woo \textit{et
al.}\cite{ref3} on bulk crystals of BCMO. We identify the peak in
magnetization to T$_{CO}$. However, this temperature is
significantly lower than the T$_{CO}$ of bulk samples ( $\approx$
330 K). \cite{ref3} Also, unlike the case of bulk samples, the
magnetization of these thin films below T$_{CO}$ does not show a
clear signature of antiferromagnetic ordering. The inset of Fig. 4
shows the M-H curve of the same sample measured at 200 K. The
dependence of magnetization on field consists of a linear and a
non-linear component. The spontaneous magnetization deduced by
extrapolation of the linear component to H = 0 comes out to be
$\approx$ 0.1 $\mu_{B}$ per Mn ion. Since in a fully aligned
ferromagnetic state this moment should be 3.4 $\mu$$_{B}$ per Mn
ion, the small value of spontaneous moment suggests the presence
of ferromagnetic correlation at T $<$ T$_{CO}$. In Fig. 5 the
temperature dependence of magnetization (M-T) of the x = 0.6
sample deposited at T$_{D}$ = 750 $^{O}$C and \opps = 170 mTorr is
shown. A comparison of these data with Fig. 4 shows a clear shift
of the peak in magnetization to lower temperatures ($\approx$ 260
K) when the O$_{2}$ partial pressure is reduced. This feature in
M-T curve becomes a mere hump centered at $\approx$ 250 K when the
O$_{2}$ partial pressure is lowered further to 70 mTorr (see Fig.
6). An interesting feature of the magnetization curves of samples
deposited at reduced O$_{2}$ partial pressures is the splitting of
the ZFC and the FC branches of the M-T curve at lower
temperatures. This observation suggests the existence of
ferromagnetically ordered clusters whose magnetization is blocked
below a critical temperature.  In the inset of Fig. 6 we show the
magnetization data for the x = 0.6 sample deposited at 750 $^{O}$C
and 650 mTorr. A cusp in the magnetization at $\approx$ 285 K
marks the T$_{CO}$. This is followed by a weak signature of
T$_{N}$ around 110 K. The cusp-like feature in magnetization,
which has been assigned to T$_{CO}$ in single crystal and bulk
samples, is pronounced in films deposited at high \opps suggests a
direct correlation between oxygen concentration in the films and
charge ordering. In Fig. 7, the result of M vs T measurement at
0.5 tesla in-plane field on a x = 0.6 sample grown at T$_{D}$ =
750 $^{O}$C and 350 mTorr \opps on (001) STO substrate is shown.
Two broad but distinct maxima in the M vs T plot observed at
$\approx$ 300 K and $\approx$ 120 K can be identified with
T$_{CO}$ and T$_{N}$ following the results on bulk samples where
these temperatures are 330 K and 125 K respectively. \cite{ref3} A
marked difference in the ZFC and FC curves is also observed below
$\approx$ 50 K. From the XRD analysis it was found that \textit{c}
for the x = 0.6 sample grown at 750$^{O}$C  on LAO at 650 mTorr
and STO at 350 mTorr were found to be 3.76 {\AA} and 3.75 {\AA}
respectively. The similarity of the magnetization curve and the
closeness of the \textit{c} value seen here indicates that the
lattice strain plays a decisive role in determining the sample
properties. The inset of Fig. 7 shows M-H measurements at 10 K and
200 K. The magnetic moment at both these temperatures is
non-linear at lower values of the applied field. The non-linear
component saturates to values which correspond to $\approx$ 0.11
$\mu$$_{B}$ and $\approx$ 0.17 $\mu$$_{B}$ moment per Mn ion at
200 K and 10 K respectively. Consistent with the behavior of ZFC
and FC magnetization at T $\approx$ 50 K, the M-H curve at 10 K
also show a hysteresis. The extremely small value of the
non-linear magnetization is consistent with measurements on bulk
samples which shows AF ordering in this temperature range. The
presence of hysteresis suggests that the bulk anti-ferromagnetic
state also has a ferromagnetic component.
 This kind of spin clustering and glassy
behavior has been observed in bulk samples with x =
0.875.\cite{ref7} A comparison of the c-axis lattice parameter
(\textit{c}) and T$_{CO}$ deduced from XRD and magnetization
measurements respectively reveal an inverse correlation between
\textit{c} and T$_{CO}$. This is shown in Fig. 8 for the x = 0.6
films deposited at 750$^{O}$C on LAO at different pO$_{2}$ and on
STO at 350 mTorr. The variation of T$_{CO}$ with \textit{c} is
linear with a slope of $\approx$ -350 K/{\AA}. Since the out of
plane lattice parameter increases with inplane compressive strain,
 this observation directly shows a compressive strain induced
suppression of the charge ordering temperature.

\subsection{Linear and non-linear electrical transport}
In Fig. 9 we show the temperature dependence of the electrical
resistivity ($\rho$(T)) for the x = 0.6 samples deposited at
different temperatures but at the same \opps of 350 mTorr. The
$\rho$(T) plots of samples deposited at T$_{D}$ $\approx$ 750
$^{O}$C show a rapid increase in the resistivity below a critical
temperature T$^{*}$ ($\approx$ 260 K) . The presence of this
feature is enhanced if we plot the differential resistivity
(d$\rho$/dT) vs T as shown in the inset for the sample deposited
at T$_{D}$ $\approx$ 750 $^{O}$C. Since the magnetization of these
samples (Fig. 4) also goes through a peak in the vicinity of this
temperature, we identify it with the charge ordering temperature
T$_{CO}$. The sudden rise in resistivity below the T$_{CO}$ is due
to the suppression of carrier hopping which is responsible for
charge transport in the charge disordered paramagnetic state
existing at T $>$ T$_{CO}$. The charge ordered state in this
system is quite robust as external field of strength $\leq$ 5
tesla does not produce any change in the resistivity. In Fig. 10
we show the variation of resistivity of the samples with different
composition but grown at the same T$_{D}$ ( = 750 $^{O}$C) and
\opps (350 mTorr). A distinct step in the resistivity that
corresponds to T$_{CO}$ is seen only for the x = 0.6 sample. This
observation is consistent with the data on bulk samples
\cite{ref3}, where the sharpness of the step in $\rho$(T) at
T$_{CO}$ diminishes as the Ca concentration is increased. In Fig.
11 the temperature dependence of the resistivity of the x = 0.6
sample deposited at T$_{D}$ = 750 $^{O}$C but at different \opps
is shown. Good quality films showing a well defined step at
T$_{CO}$, are found to grow in the \opps regime of 170 to 350
mTorr. From these studies we can conclude that best conditions for
the growth of charge ordered BCMO films correspond to T$_{D}$ =
750 $^{O}$C - 800 $^{O}$C and \opps $\approx$ 170 - 350 mTorr. In
Fig. 12 we show the temperature dependence of resistivity of the x
= 0.6 sample grown on STO at 750$^{O}$C and 350 mTorr \opps. In
this sample the step in the resistivity at T$_{CO}$ is not very
pronounced as compared to the sample grown on LAO. The inset shows
the variation of the differential resistivity as a function of
temperature. The kink in this graph at $\approx$ 310 K corresponds
to the T$_{CO}$, which is close to the value obtained from the
magnetization measurements on this sample. The in-plane
resistivity of the film on STO is approximately one order
magnitude larger than that of the sample grown on LAO. This
observation can be understood on the basis of increased tensile
strain which reduces the Mn-O-Mn orbital overlap in the plane of
the film thereby favoring electron localization.

We have analyzed the temperature dependence of electrical
resistivity of the x = 0.6 samples at T $<$ T$_{CO}$ in some
detail. A thermally activated transport due to excitation of
carriers across a charge order gap should lead to a resistivity of
the form $\rho$ = $\rho_{0}$ exp[ $\Delta$E/K$_{B}$T]. A plot of
ln $\rho$ vs 1/T (not shown) showing a continuously changing slope
of this curve as temperature is lowered below T$_{CO}$ is
indicative of a temperature dependent activation energy (
$\Delta$E(T)). For the sample deposited at 750 $^{O}$C for
example, the activation energy changes from 0.2 eV to 0.1 eV as
the temperature is lowered from T$_{1}$ (= 225 K) to T$_{2}$ (=
150 K). At T $<$ T$_{2}$, the activation energy remains nearly
constant. The simplest scenario that can account for such a
dependence of $\Delta$E is presented by Mott variable range
hopping (VRH) process. \cite{ref20} The VRH transport in a Fermi
glass \cite{ref20} leads to a resistivity of the type $\rho$ (T) =
$\rho$$_{0}$ exp[{(T$_{0}$/T)$^{1/4}$], where $\rho$$_{0}$ is the
preexponential factor and T$_{0}$ is related to the density of
localized states at the Fermi energy N(E$_{F}$) through the
relation K$_{B}$T$_{0}$ = 18$\alpha$$^{3}$/N(E$_{F}$). In Fig. 13
we show plots of ln($\rho$) vs T$^{1/4}$. The VRH formalism was
found to be  in better agreement with the behavior of $\rho$(T) at
T $<$ T$_{CO}$. The slopes of these curves yield K$_{B}$T$_{0}$ of
11 KeV, 16 KeV and 14 KeV for the sample deposited at 700, 750 and
800 $^{O}$C respectively. The value of K$_{B}$T$_{0}$ for the
sample grown on STO was found to be  19 KeV. A calculation of the
localization length a = (1/$\alpha$) from the K$_{B}$T$_{0}$
requires a knowledge of the N(E$_{F}$), which can be estimated
from the measurement of the electronic specific heat or Pauli spin
susceptibility. While specific heat has been measured for
less-than-half-filled manganites \cite{ref25,ref27,ref28}, such
studies are lacking in the case of BCMO. The electronic
susceptibility is also overwhelmed by the contribution from the
localized t$_{2g}$ spins at the Mn sites. In view of these
limitations, the observation of VRH only signifies the presence of
disorder in the medium.

It is experimentally well established that the charge ordered
manganites with less-than-half-filling (LTHF) show fascinating
effects of electric field on charge transport.
\cite{ref21,ref22,ref31} In order to see if similar effects also
exists in this more-than-half-filled (MTHF) material, we have
carried out the measurement of current-voltage characteristics of
the best film of x = 0.6. All measurements were done in the
constant voltage mode. The variation of the circuit current (I) as
a function of applied circuit voltage (V) was measured at
different temperatures by calculating the drop across a 1
k$\Omega$ metal film resistor connected in series with the sample.
Holding the temperature constant, the applied voltage (V) was
swept up to a maximum value and then swept down at the same rate.
In Fig. 14 (a) results of I-V measurements at different
temperatures are shown. When the applied voltage across the sample
reaches a threshold value V$^*$, the sample makes a transition
from the insulating state to a low resistive state as seen from
the sudden jump in the current I. On further increasing the
applied voltage, the current in the circuit is limited by the
standard resistor in series with the sample. The sample switches
back to its insulating state on decreasing the V but at a value
much lower than the V$^*$. This results in a hysteretic I-V curve.
Such hysteretic behavior has been seen abundantly in
less-than-half-filled manganites. \cite{ref21,ref22,ref23}
However, unlike the case of LTHF manganites where the hysteretic
response disappears on measuring the I-V again at the same
temperature, here repeated measurements at the same temperature
produce a hysteretic response. The lifetime of the current driven
conducting state is only a few seconds after switching off the
electric field. This observation suggests that the metallic state
is perhaps induced by sample heating. We have repeated these
measurements on several samples of different bridge dimensions and
the switching is observed in all cases. The switching voltage
depends upon the bridge dimension, for smaller length of the
bridge, the current switching occurs at lower voltages. In Fig. 14
(b)  the circuit current (I) is plotted as a function of the
voltage (V) across the sample at 120 K. The bridge width was 50
$\mu$m. These measurements were done simultaneously with the
measurements shown in  Fig. 14(a). Similar behavior of I-V curves
has  been observed in epitaxial thin films of
Sm$_{0.5}$Sr$_{0.5}$MnO$_{3}$ by Oshima \textit{et
al}.\cite{ref34} The negative differential resistance exhibited by
the sample in the electric field induced low resistive state
strongly suggests that the current path is filamentary in
nature.\cite{ref18,ref22}Similar kind of switching behavior was
also seen for the sample grown on STO. In order to address this
issue in detail, the I-V measurements were performed by varying
the width (\tw) of the voltage pulse and the time delay (\td)
between two successive pulses. It was found that as the time
interval between pulses is reduced (\tds $\approx$ 1 sec) the
current switching occurs gradually. It was also observed that when
this interval was increased (\tds $\approx$ 12 sec), there was no
current switching at all, even for the highest V, as shown in Fig.
15. This observation suggests that the current-switching
phenomenon in this MTHF manganite is primarily thermal in nature.
However, if the measurements are repeated at a particular
temperature keeping \tds same for all the scans and by making sure
that sample reaches thermal equilibrium after each scan, it is
found that for later scans current switching occurs at a lower
value of the voltage (V*) as seen in Fig. 16. However, the V$^*$
does not keep decreasing with the increasing number of scans, but
reaches a constant value $V^{*}_{C}$ after few scans, typically
five. Here it needs to be emphasized that if the current switching
was purely a heating effect, the same nature of the I-V curves
would appear for all scans, until and unless enough thermalization
time is not given, which is not the case here. Interestingly, the
virgin behavior of I-V curve at a particular temperature T $<$
T$_{CO}$ is recovered ($V^{*}$ $>$ $V^{*}_{C}$) after cycling the
sample to T $>$ T$_{CO}$. From the data of Fig. 14, 15, and 16 it
can be argued that the bias dependence of the current transport in
these films is a compound effect of simple heating, which is
reversible, and a fractional permanent change in the topology of
the CO state which creates patches of conducting regions. These
conducting domains appear to grow in size during successive I-V
cycles till saturation. This causes the circuit voltage, at the
point of switching to reach the constant value ($V^{*}_{C}$) after
a few scans. It is to be noted that while measuring I-V
characteristics at different temperatures the lowest temperature
measurement was done first, and repeated scans were made to
establish the constant voltage ($V^{*}_{C}$) and then only the
temperature was increased. Charge ordering becomes more and more
robust as the temperature is decreased below T$_{CO}$. This causes
the $V^{*}_{C}$ to increase with decreasing temperature as shown
in the inset of Fig. 16. It should be recalled that in
less-than-half-filled manganites (x $<$ 0.5) of marginal bandwidth
\cite{ref29,ref30,ref33}, the energy difference between the CDFM
and CO states is small, and the system can switch from CO to CDFM
states on application of small perturbations. For
more-than-half-filled manganites (x $>$ 0.5) such as the
Bi$_{0.4}$Ca$_{0.6}$MnO$_{3}$ in particular, there is no evidence
for the formation of a CDFM state. The observed response here is a
mixture of thermal and electronic effects.

\subsection{Photoresponse measurements}
We have also studied the stability of the charge ordered state
under illumination with 355 nm pulsed radiation. These
measurements were done in isothermal mode by biasing the circuit
at a fixed voltage V $\leq$ $V^{*}_{C}$, where $V^{*}_{C}$ is the
circuit voltage at which the circuit current shows the distinct
step. The oscilloscope was connected directly across the sample to
record the conductivity profile, in terms of the voltage drop, in
real time. These measurements were done at temperatures where the
sample resistance was orders of magnitude less than the input
impedance of the scope. Fig. 17 shows the transient conducting
state (TCS) induced by the 6 ns laser pulse at 120 K where the
sample is initially insulating [V$<$ $V^{*}_{C}$]. The photon flux
was $\approx$ 10$^{15}$ photon/mm$^{2}$ which corresponds to an
energy density of 0.16 J/cm$^{2}$. The profile has a leading edge
that follows the pulse shape. Here the sample resistance drops
precipitously. This is followed by a tailing effect where the
sample recovers its original insulating state. This recovery can
be modeled by a bounded exponential growth function of the form
V(t) = V$_{0}$ + V$_{1}$(1$\--$e$^{ -t/\tau}$). The value of
$\tau$, which signifies the time constant was found to be
$\approx$ 2.4 $\mu$s. In Fig. 18 we show the dependence of the
photoresponse on incident power for a given bias voltage. A power
of 90$\%$ in the figure corresponds to energy density of 0.16
J/cm$^{2}$. Two features of these data are noteworthy. First, the
response has a damped oscillatory behavior over short time scale
of the recovery process ($\approx$ 2$ \mu$s). A similar behavior
has also been observed by Miyano \textit{et al.}.\cite{ref8} in
the case of Pr$_{0.7}$Ca$_{0.3}$MnO$_{3}$ crystals. This is an
effect of the sudden voltage drop across the sample caused by the
large photocurrent. Secondly the photoresponse is non-linear in
power as shown in the inset of Fig. 18. The threshold power
required for an observable change is $\approx$ 40$\%$. The
fundamental difference between the photoresponse of BCMO as
studied here and those of less-than-half-filled charge ordered
manganites such as Pr$_{0.7}$Ca$_{0.3}$MnO$_{3}$ \cite{ref10} and
Pr$_{1-X}$(Ca$_{Y}$Sr$_{1-Y}$)$_{X}$MnO$_{3}$ \cite{ref11} is the
behavior of resistance on removal of the single shot photon field
of duration $\approx$ 6 ns. In the present case the system
recovers its insulating state on removal of the photon field. In
PCMO, however, the photoinduced metallic state persists as long as
the bias voltage is present. The sample goes back to the
insulating state on removal of the bias. If the strength of the CO
state in PCMO is weakened through bandwidth control, the
photoinduced metallic state survives even on removal of the bias,
as seen in the case of
Pr$_{0.55}$(Ca$_{1-Y}$Sr$_{Y}$)$_{0.45}$MnO$_{3}$ (0.2 $\leq$ y
$\leq$ 0.4).\cite{ref11} The absence of a persistent metallic
state in these BCMO films after the single-shot photoexposure is
an indication of the high stability of the CO-state. This
inference is consistent with the results of electric and magnetic
field induced changes in charge transport, both of which reveal a
robust CO-state. An interesting aspect of photoillumination
emerges when the sample is driven to the conducting state by
applying voltages more than critical switching field. A typical
response of the sample in the field-driven metallic state is shown
in Fig. 19. Unlike the previous case, here both the leading and
the trailing edges of the conductivity profile follow the shape of
the laser pulse. The characteristic recovery time $\tau$ in this
case is only $\approx$ 60 ns. This observation further suggests
that the long recovery processes (several microseconds) seen in
Fig. 17 and 18 is due to the reappearance of COI domains which had
melted under photoillumination. Finally we compare our results of
pulsed laser irradiation of BCMO films with the work on
photoillumination of BCMO crystal with a 488 nm CW laser.
\cite{ref6} The irreversible changes seen in the refractive index
of those crystals by Smolyaninov \textit{et al.}\cite{ref6} were
realized at a much higher photon flux ($\approx$ 10$^{24}$
photon/mm$^{2}$). We believe the reversibility seen in our films
is due to the lower photon flux used in these experiments. The
CO-state in films is also likely to get pinned by growth related
strain and defects.

\subsection{CONCLUSIONS}
In conclusion, epitaxial thin films of \bcmo (x = 0.6, 0.67, 0.76)
were grown on LAO (100) and STO (100) substrates using pulsed
laser deposition. The out-of-plane lattice parameter of the films
is a sensitive function of the $T_{D}$ and \opps and the lattice
parameter of the substrate, and plays a key role in setting the
temperature scale for charge ordering. Magnetization measurements
on Bi$_{0.4}$Ca$_{0.6}$MnO$_{3}$ grown on LAO under optimal
condition reveal a  $T_{CO}$ of $\approx$ 275 K  and no signatures
of $T_{N}$ unlike in bulk samples for which $T_{CO}$ $\approx$ 330
K and $T_{N}$ $\approx$ 125 K. However,
Bi$_{0.4}$Ca$_{0.6}$MnO$_{3}$ grown on STO exhibits $T_{CO}$
$\approx$ 300 K and $T_{N}$ $\approx$ 120 K apart from a small
ferromagnetic phase below 50 K. This difference can be attributed
to the fact that BCMO grown on STO experiences an in-plane tensile
strain owing to the difference in their lattice parameters.
Electrical transport in the charge-ordered state has signatures of
variable range hopping in a three dimensional disordered system,
which presumably results from the A site ionic size and positional
disorder in the perovskite lattice. I-V measurements on the
Bi$_{0.4}$Ca$_{0.6}$MnO$_{3}$ exhibited hysteretic current
switching with a history dependent voltage threshold which appears
to be a compound effect of joule heating and permanent changes in
the topology of the CO state. Photoresponse measurements using
nano second pulses of the third harmonic of Nd-YAG laser were
performed on  Bi$_{0.4}$Ca$_{0.6}$MnO$_{3}$  both in its
insulating state as well as the electric field induced metallic
state. The response was transient with a lifetime of $\approx$ 2.4
$\mu$s and 60 ns for the insulating and metallic states
respectively. Unlike the case of some less-than-half-filled CO
manganites of intermediate bandwidth where the photoinduced
conducting state persists on removal of the photon field, the
quick recovery seen here indicates that the CO state is quite
robust in these Bi$_{0.4}$Ca$_{0.6}$MnO$_{3}$ epitaxial films.

$\textbf{ACKNOWLEDGEMENTS}$

 This research has been supported by a grant from the Indo-French
 Centre for Promotion of Advanced Research (IFCPAR) New Delhi,
India.

\clearpage
\begin{figure}
\includegraphics [width=16cm]{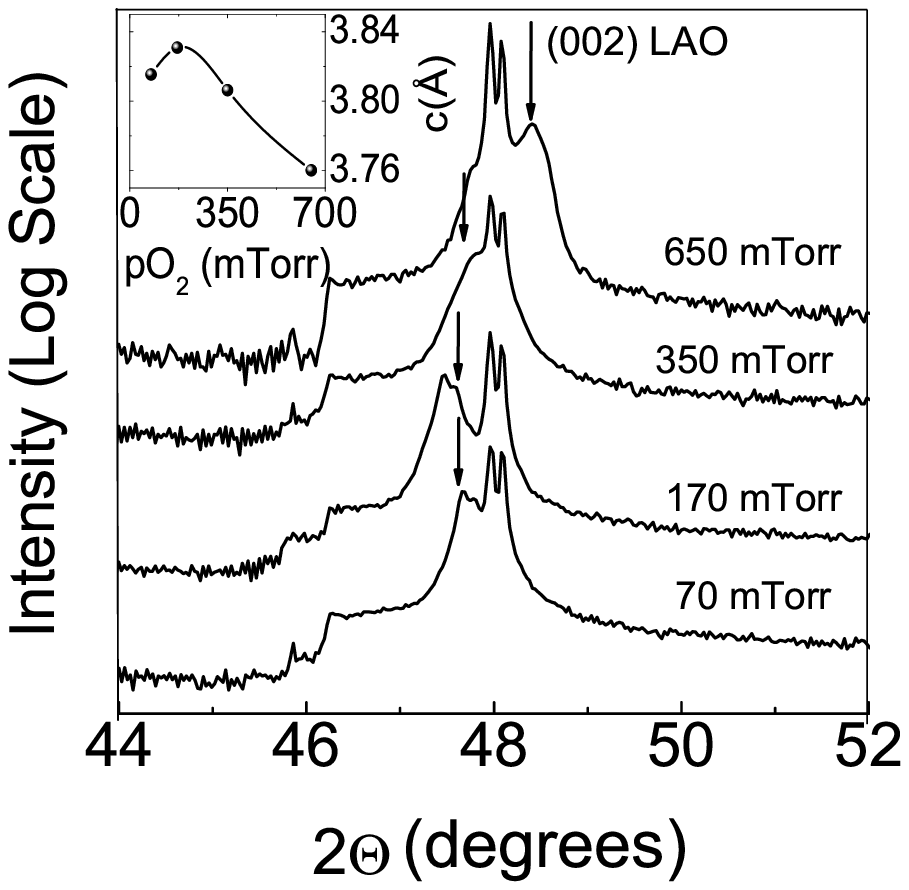}
\caption{\label{Fig1}X-ray diffraction  pattern of  \bcmo thin
films on LAO with x = 0.6 grown at 750 $^{O}$C  under different
oxygen partial pressure. The inset shows the variation of  the
out-of-plane lattice parameter (\textit{c}) with \oppsb. The
diffraction peak of the film is marked by arrow.}
\end{figure}

 \clearpage
\begin{figure}
\includegraphics [width=16cm]{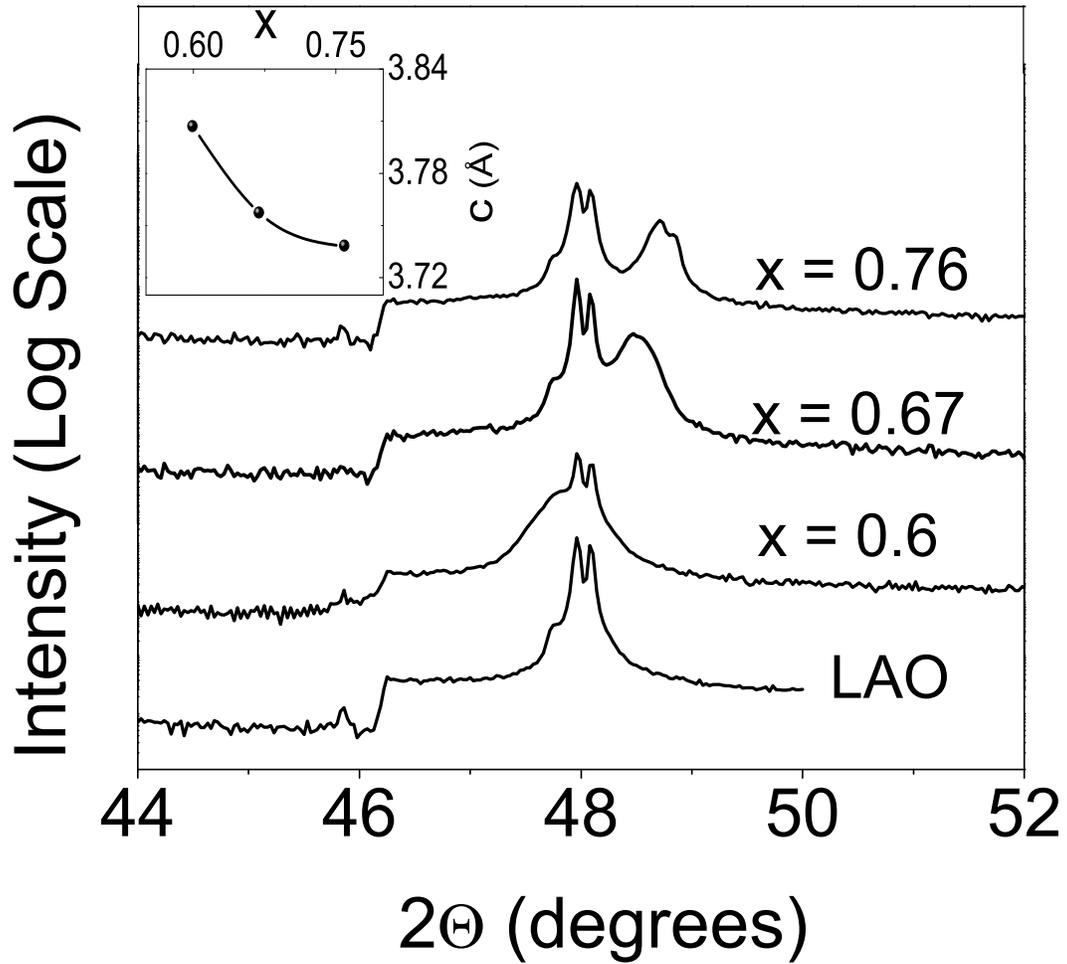}
\caption{\label{Fig2}X-ray diffraction  pattern of  \bcmo thin
films on  LAO  for different x grown at 750 $^{O}$C and oxygen
partial pressure of 350 mTorr.  The inset shows the variation of
the out-of-plane lattice parameter  with x.}
\end{figure}

\clearpage
\begin{figure}
\includegraphics [width=16cm]{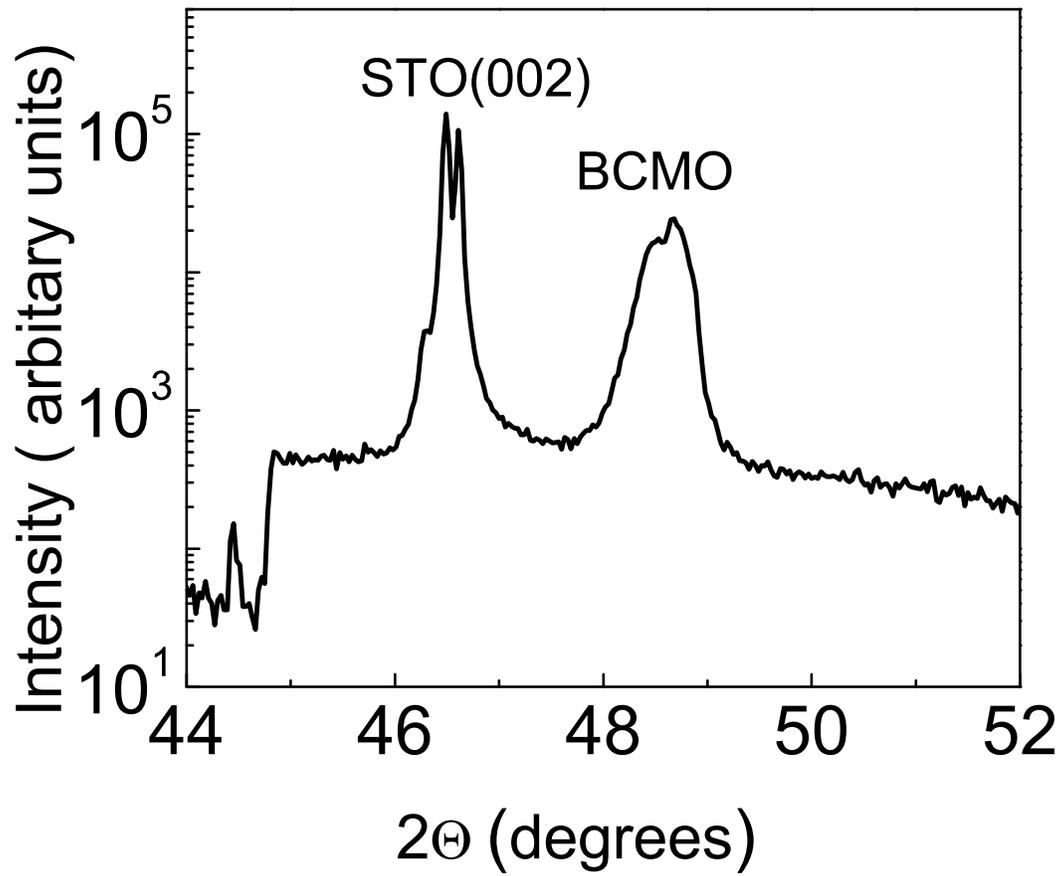}%\
\caption{\label{fig3}X-ray diffraction  pattern of  \bcmo thin films on STO  with x = 0.6 grown at  oxygen partial pressure of
350 mTorr and T$_{D}$ $\approx$ 750 $^{O}$C.}
\end{figure}

\clearpage
\begin{figure}
\includegraphics [width=16cm]{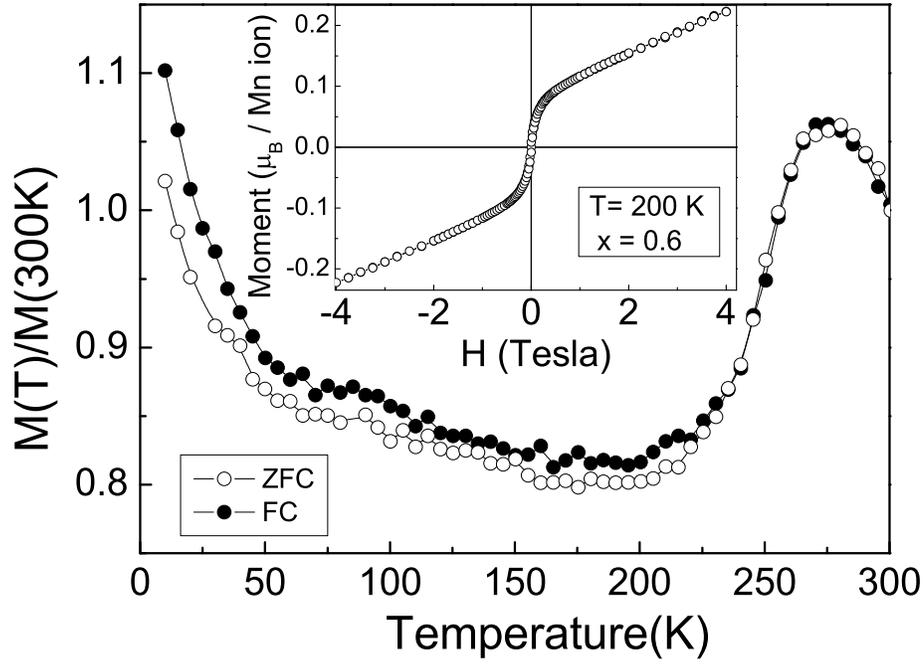}%\
\caption{\label{fig4}Field-cooled and Zero-field-cooled
magnetization vs temperature plots for
Bi$_{0.4}$Ca$_{0.6}$MnO$_{3}$ film on LAO grown at 750 $^{O}$C and
350 mTorr. The measurements were performed in an in-plane field of
0.5 T. The peak at $\approx$ 275 K corresponds to T$_{CO}$. Inset
shows the behavior of magnetization as a function of field at 200
K.}
\end{figure}

\clearpage
\begin{figure}
\includegraphics [width=16cm]{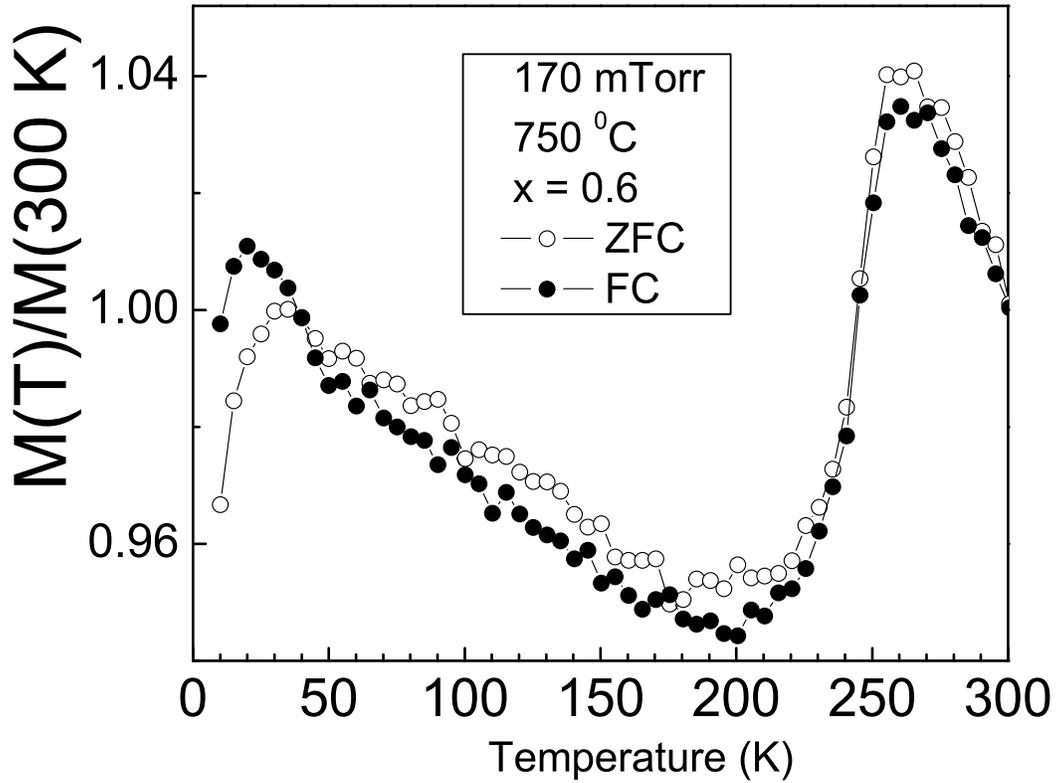}%\
\caption{\label{fig5}Field-cooled and Zero-field-cooled magnetization vs temperature plots for Bi$_{0.4}$Ca$_{0.6}$MnO$_{3}$
film on LAO grown at 750 $^{O}$C and 170 mTorr. The measurements were performed in an in-plane field of 0.5 T. The peak at
$\approx$ 260 K corresponds to T$_{CO}$.}
\end{figure}

\clearpage

\begin{figure}
\includegraphics [width=16cm]{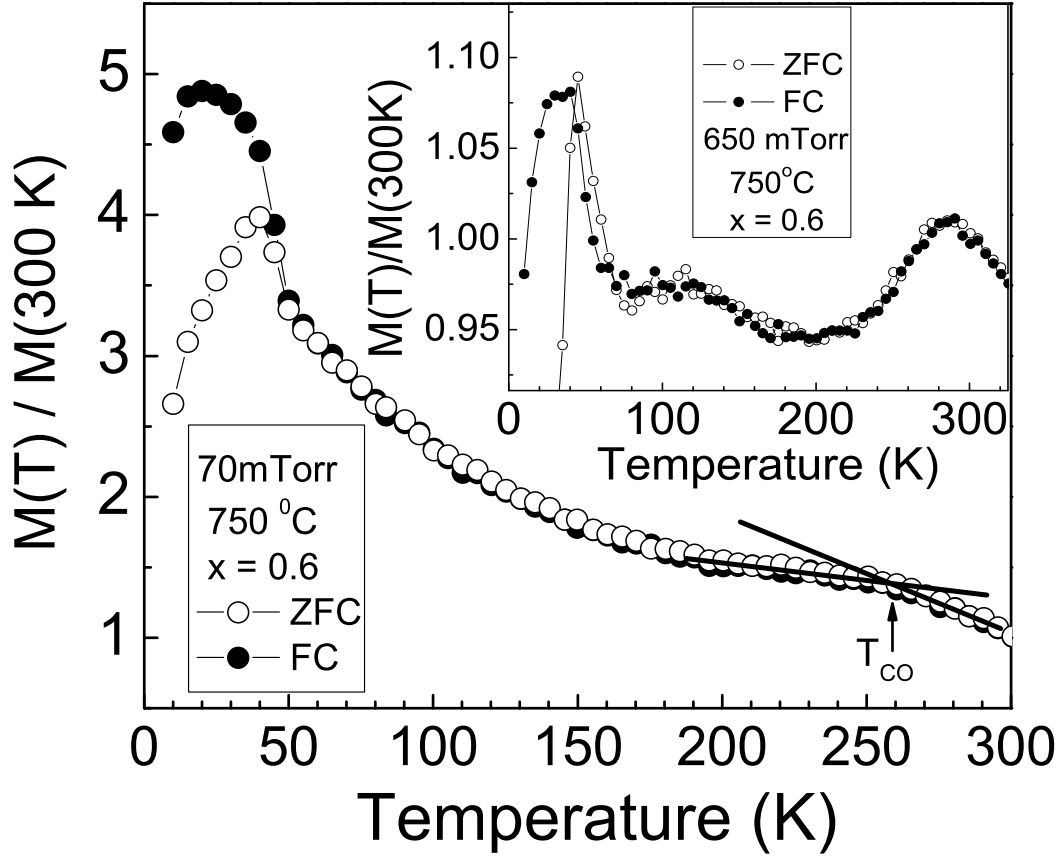}%\
\caption{\label{fig6}Field-cooled and Zero-field-cooled
magnetization vs temperature plots for
Bi$_{0.4}$Ca$_{0.6}$MnO$_{3}$ film on LAO grown at 750 $^{O}$C and
70 mTorr. The measurements were performed in an in-plane field of
0.5 T. There is no signature of T$_{CO}$ in the data. The inset
shows the magnetization measurement of the x = 0.6 sample
deposited at 750 $^{O}$C and 650 mTorr in an in-plane field of 0.5
T. The peak at $\approx$  285 K and 110 K corresponds  to T$_{CO}$
and T$_{N}$ respectively.}
\end{figure}

\clearpage
\begin{figure}
\includegraphics [width=16cm]{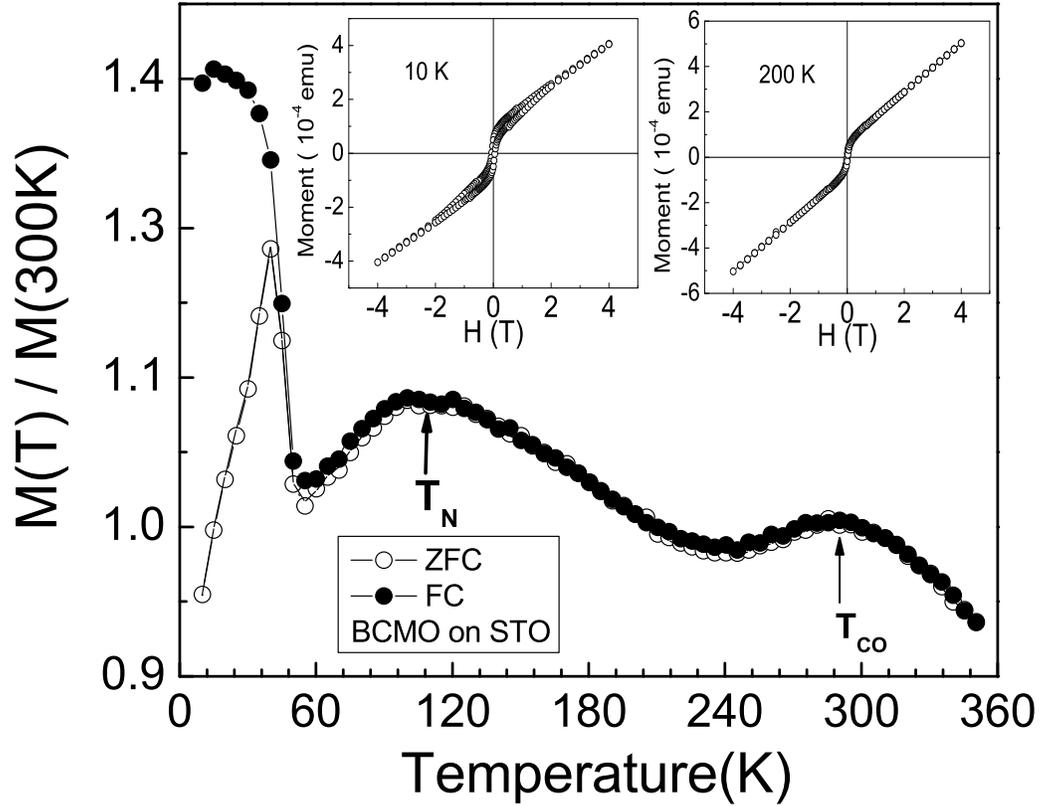}%\
\caption{\label{fig7}Field-cooled and Zero-field-cooled magnetization vs temperature plots for a Bi$_{0.4}$Ca$_{0.6}$MnO$_{3}$
film deposited on STO at 750 $^{O}$C and 350 mTorr of \opps. The applied field of 0.5 T was in the plane of the film. The peak
at $\approx$ 300 K and $\approx$ 125 K corresponds to  T$_{CO}$ and T$_{N}$ respectively. The  inset  shows M-H data at 10 K and
200 K .}
\end{figure}

\clearpage
\begin{figure}
\includegraphics [width=16cm]{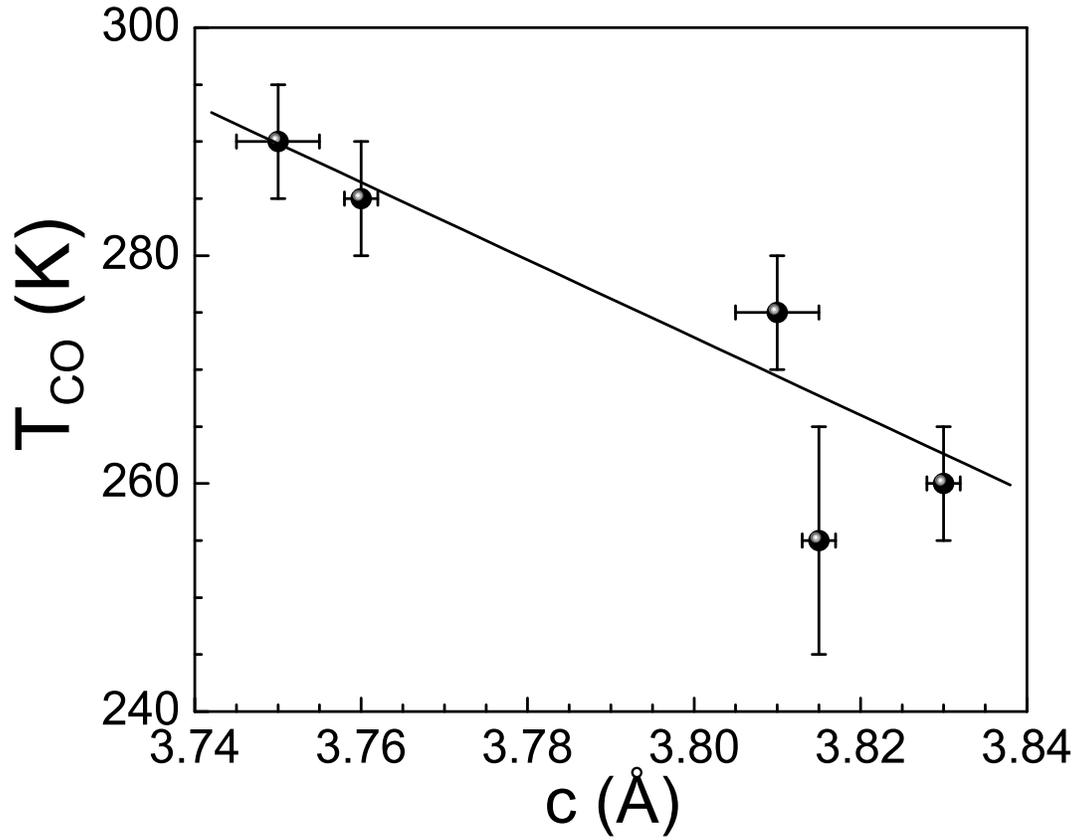}%\
\caption{\label{fig7n} The variation of T$_{CO}$  with out of
plane lattice parameter (\textit{c}) for BCMO with x = 0.6 and
deposited at
 750$^{O}$C. }
\end{figure}

\clearpage
\begin{figure}
\includegraphics [width=16cm]{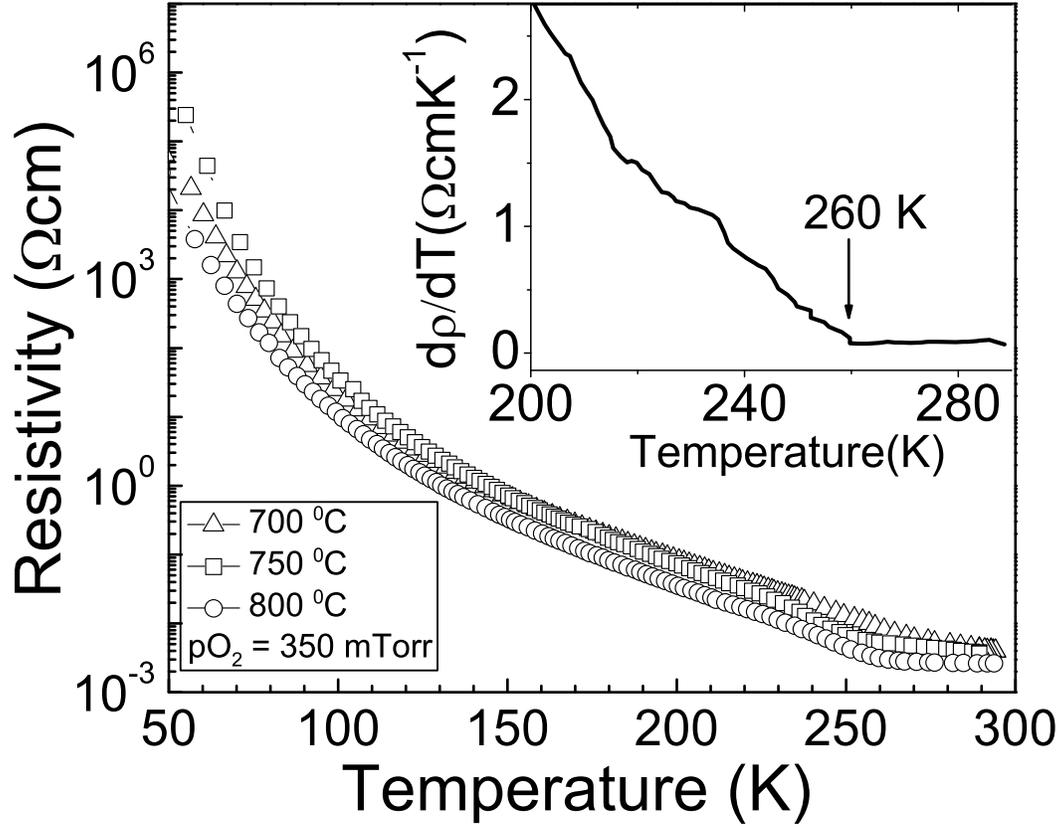}%\
\caption{\label{fig8}Electrical resistivity of
Bi$_{0.4}$Ca$_{0.6}$MnO$_{3}$ thin films deposited on LAO at a
fixed \opps (350 mTorr) but at different temperatures (T$_{D}$)
plotted as a function of temperature. A pronounced step at
$\approx$ 260 K is observed for the sample with T$_{D}$ of
$\approx$ 750 $^{O}$C. The  inset shows the differential
resistivity as a function of temperature for the same sample.}
\end{figure}

 \clearpage
\begin{figure}
\includegraphics [width=16cm]{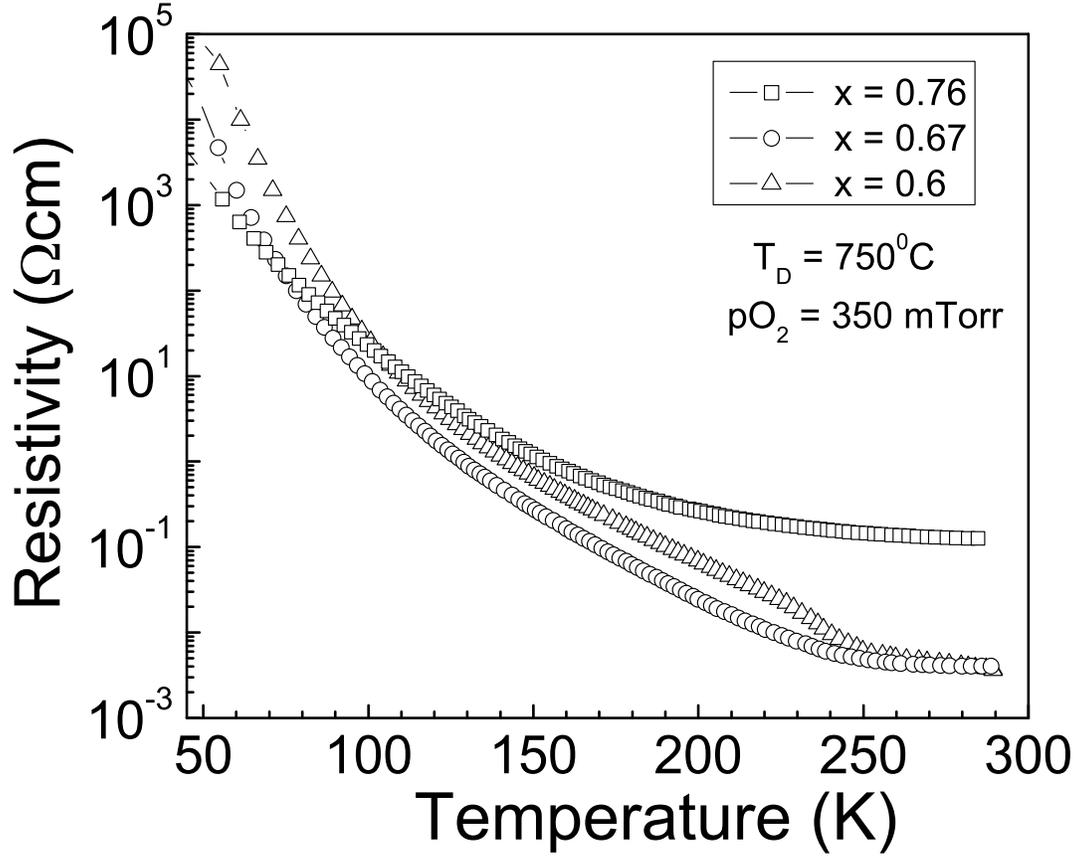}%\
\caption{\label{fig9}Electrical resistivity of \bcmo (x = 0.6,
0.67, 0.76) thin films deposited on LAO  at T$_{D}$ $\approx$ 750
$^{O}$C and \opps of 350 mTorr plotted as a function of
temperature. A pronounced step at $\approx$ 260 K is observed for
the sample with x = 0.6.}
\end{figure}

\clearpage
\begin{figure}
\includegraphics [width=16cm]{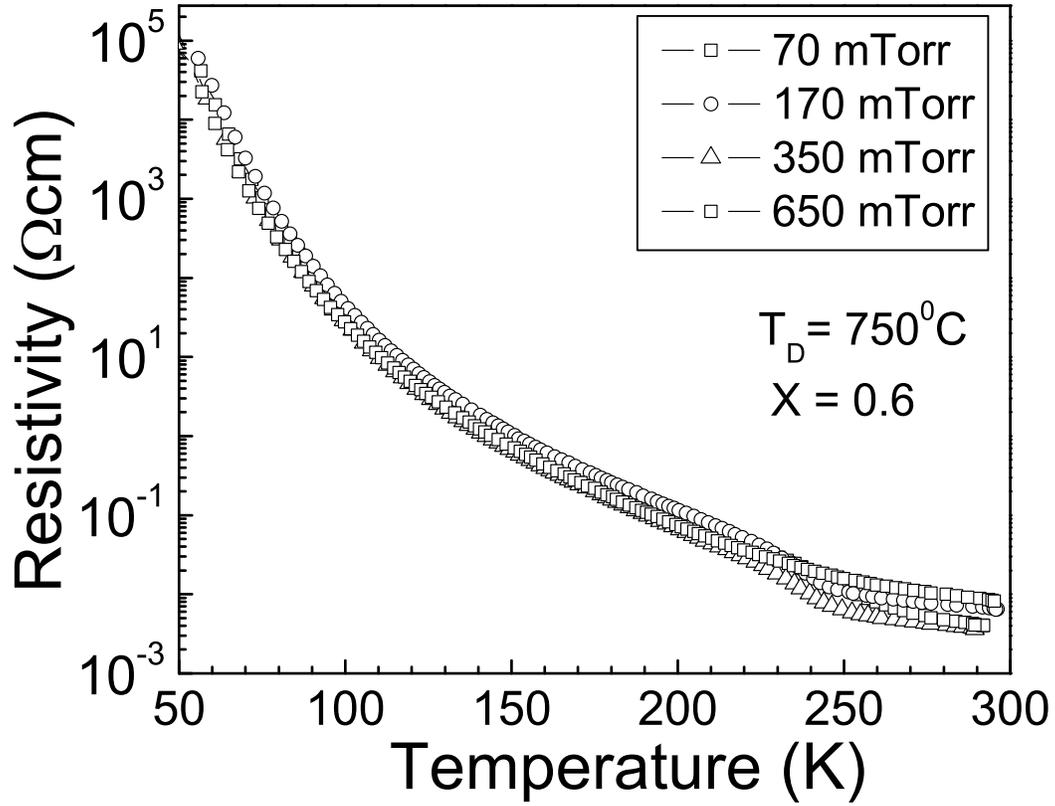}%\
\caption{\label{fig10}Electrical resistivity of
Bi$_{0.4}$Ca$_{0.6}$MnO$_{3}$ thin films deposited on LAO at
T$_{D}$ $\approx$ 750 $^{O}$C but at different oxygen partial
pressure (\opps) plotted as a function of temperature. A
pronounced step at $\approx$ 260 K is observed for the sample
deposited at 350 mTorr.}
\end{figure}

\clearpage
\begin{figure}
\includegraphics [width=16cm]{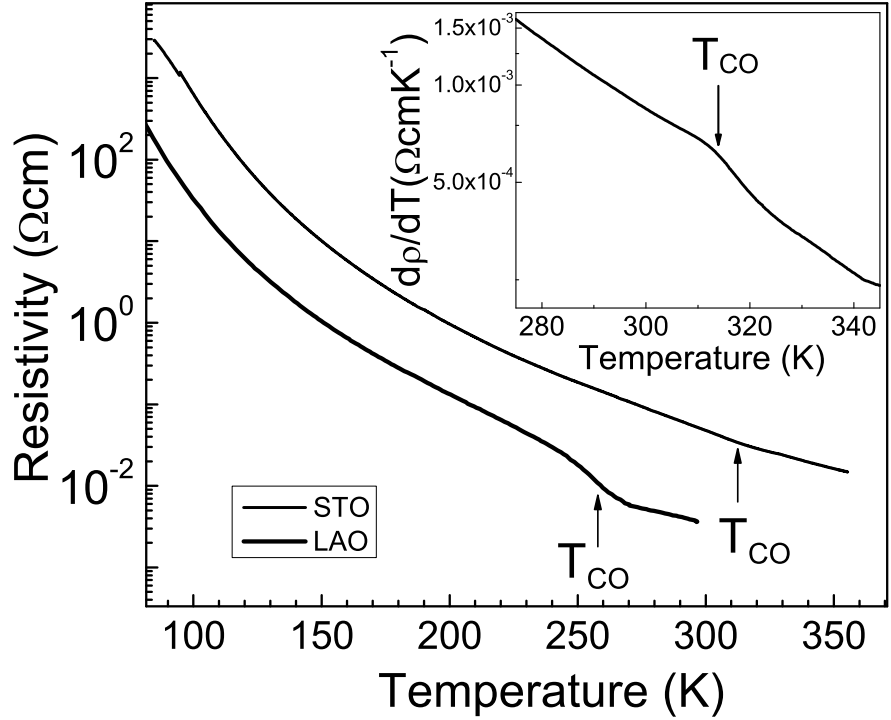}%\
\caption{\label{fig11}Electrical resistivity of
Bi$_{0.4}$Ca$_{0.6}$MnO$_{3}$ thin films deposited on STO  and LAO
at 750$^{O}$C and \opps of 350 mTorr  plotted as a function of
temperature.  The  inset shows the differential resistivity as a
function of temperature for the  sample on STO. The kink at
$\approx$ 310 K corresponds to T$_{CO}$. }
\end{figure}

\clearpage
\begin{figure}
\includegraphics [width=16cm]{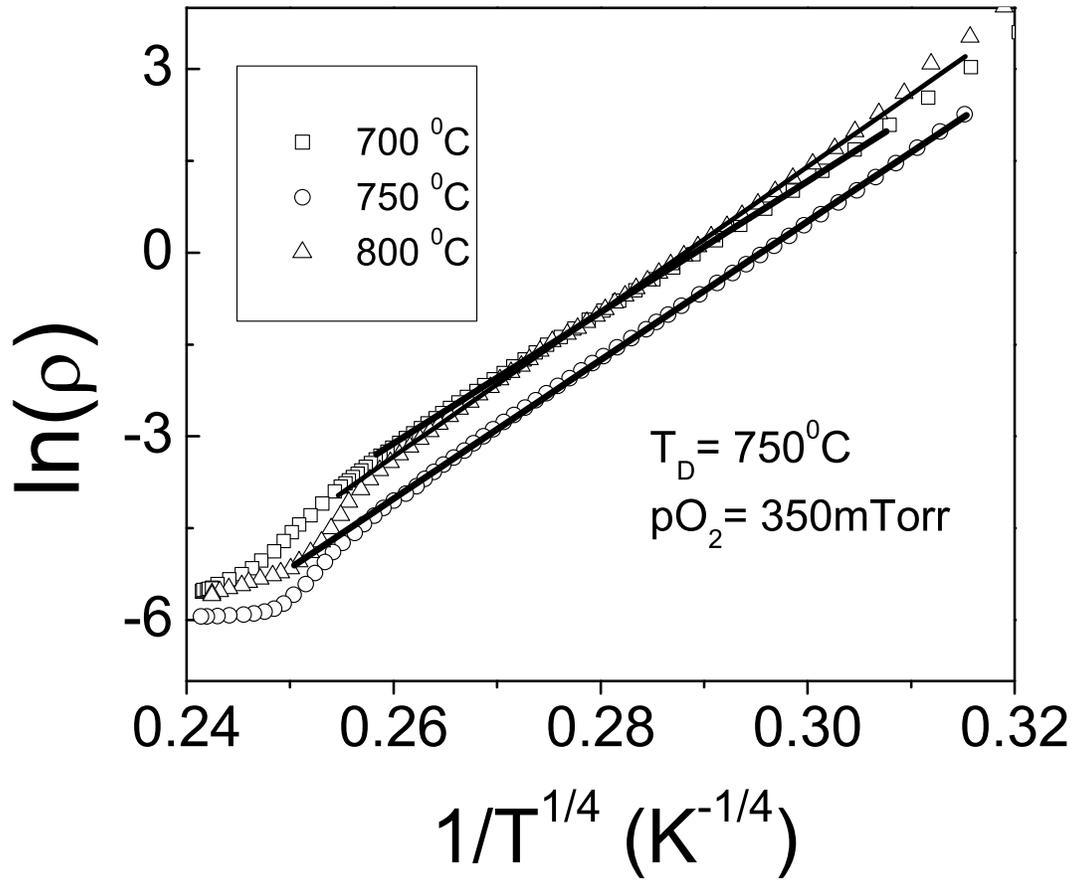}%\
\caption{\label{Fig12}Ln ($\rho$(T)) vs  T$^{-1/4}$ plots for the sample with x = 0.6 grown at fixed \opps (350 mTorr) but
different deposition temperature of 700 $^{O}$C, 750 $^{O}$C and 800 $^{O}$C. The solid lines show  fitting to the variable
range hopping formula $\rho$(T) =$\rho$$_{0}$ exp [T$_{0}$/T]$^{1/4}$ .}
\end{figure}

\clearpage
\begin{figure}
\includegraphics [width=16cm]{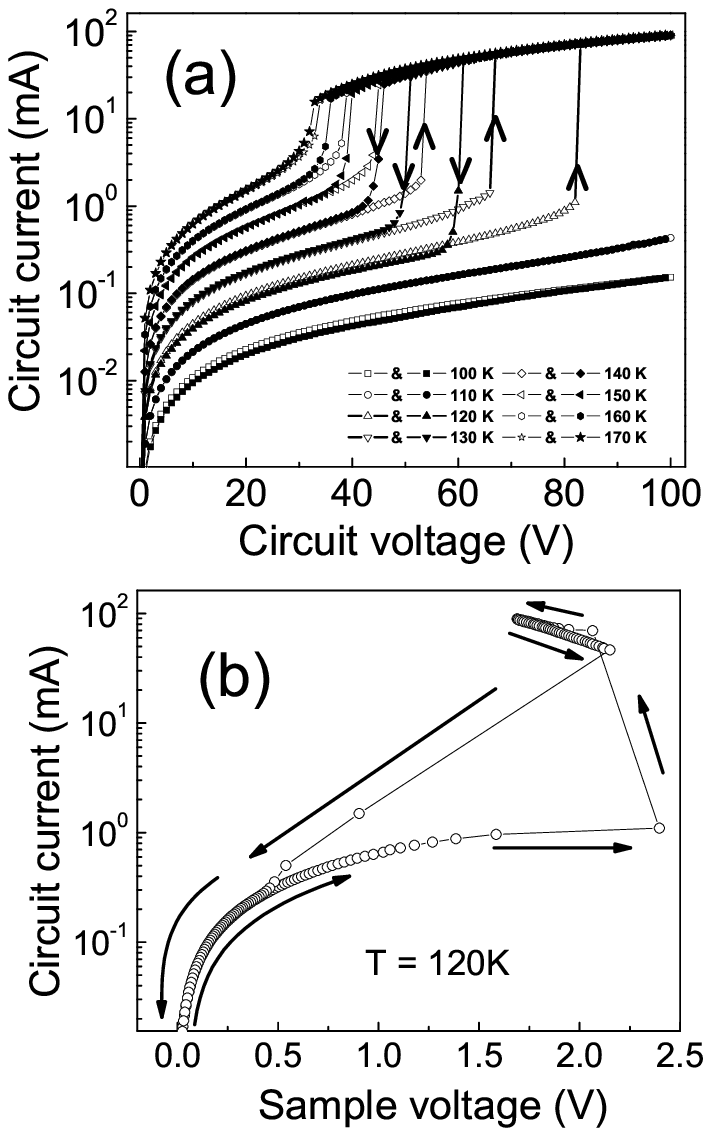}%\
\caption{\label{fig13}(a)I-V characteristics of the circuit
comprising of the BCMO film (x = 0.6) deposited at 750 $^{O}$C and
a 1 k$\Omega$ resistor in series. The measurements have been done
at different temperatures. The open and filled symbols correspond
to the increasing and decreasing branch of the voltage cycle
respectively. (b)I-V characteristics of the same BCMO film at
120K. This measurement was done simultaneously with the above
measurement.}
\end{figure}

\clearpage
\begin{figure}
\includegraphics [width=16cm]{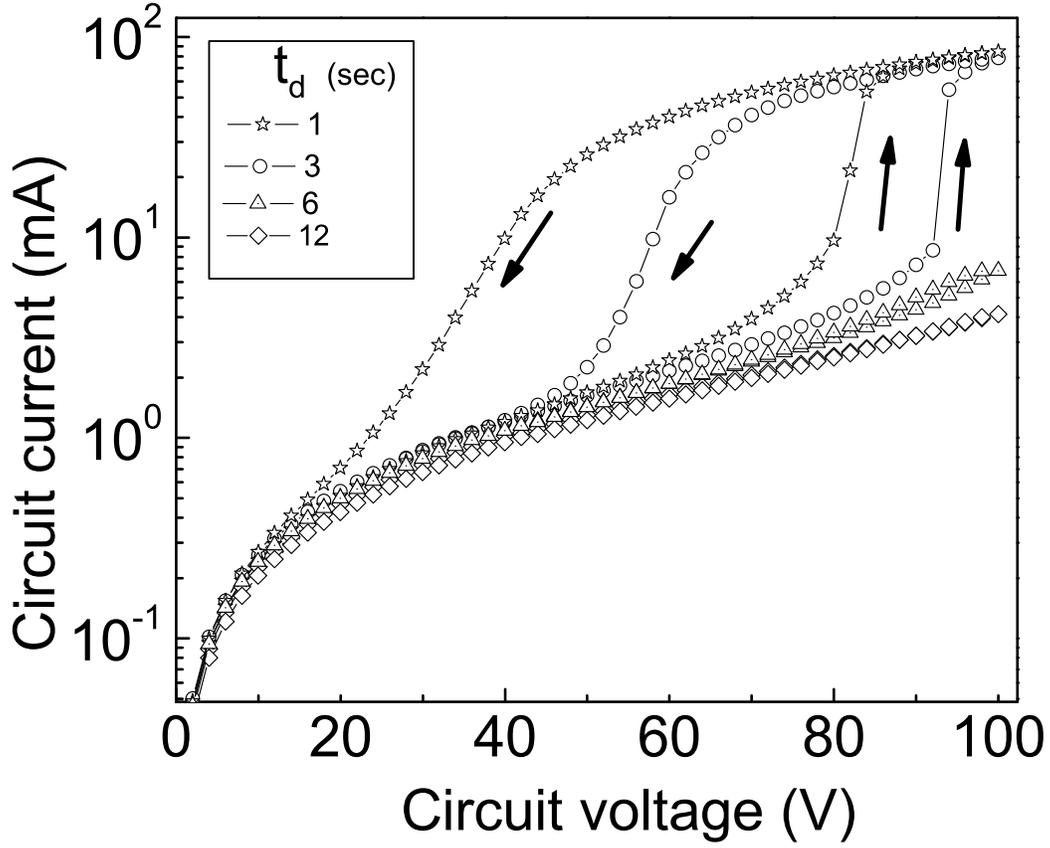}%\
\caption{\label{fig14}The variation of circuit current (I) as a function of applied voltage (V) at 120 K. The measurements were
done by increasing the voltage in a stepped manner. The parameter \tds  is the time during which the voltage pulse was off. The
increasing and the decreasing branches of the I-V are marked.}
\end{figure}

 \clearpage
\begin{figure}
\includegraphics [width=16cm]{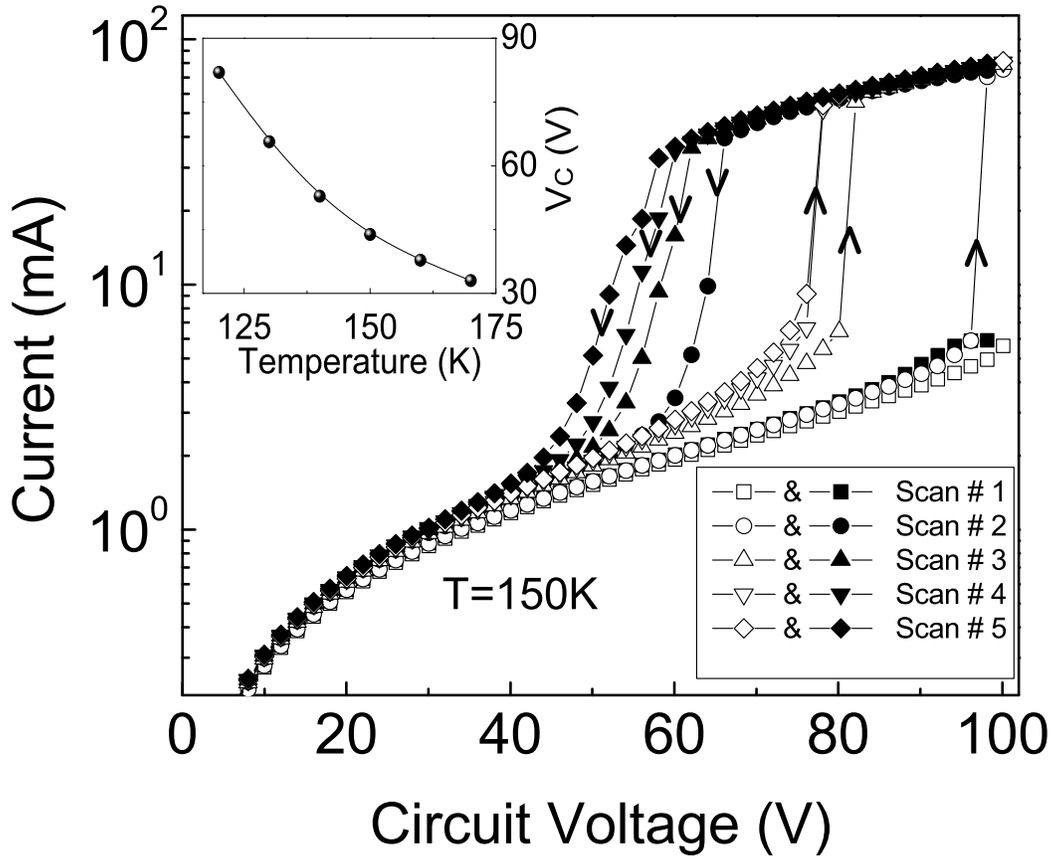}%\
\caption{\label{fig15}The variation of circuit current (I) as a function of applied  (V) at 150 K for successive scans. The open
symbols correspond to ramping the voltage to higher values and the closed one corresponds to the reverse ramping. The inset
shows the variation of the critical  applied circuit voltage $V^{*}_{C}$ (at which the sample goes to a conducting state) as a
function of temperature.}
\end{figure}

\clearpage
\begin{figure}
\includegraphics [width=16cm]{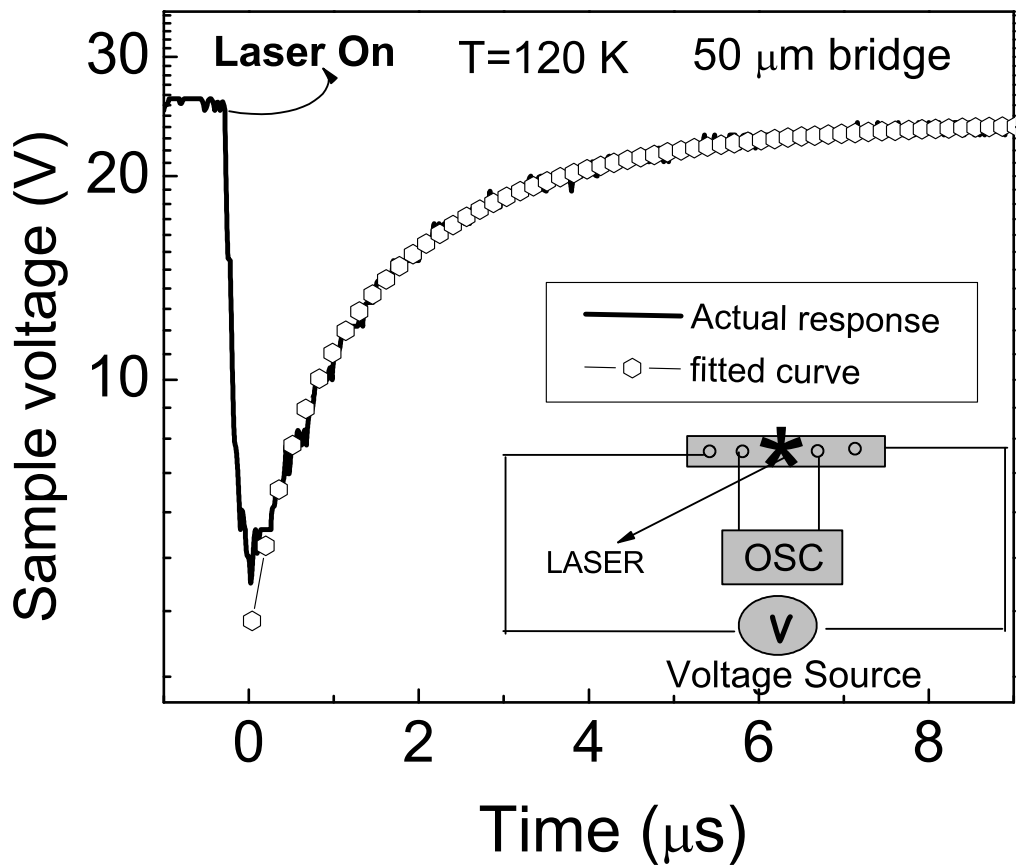}%\
\caption{\label{fig16}The variation of the sample voltage in real time as a result of photoillumination at 120 K where the
sample is initially in the charge ordered insulating state. The sample regains its original insulating state in about 2.4 $\mu$s
after switching off the 6 ns photon pulse. The basic circuit diagram is shown in the bottom right corner. }
\end{figure}

\clearpage
\begin{figure}
\includegraphics [width=16cm]{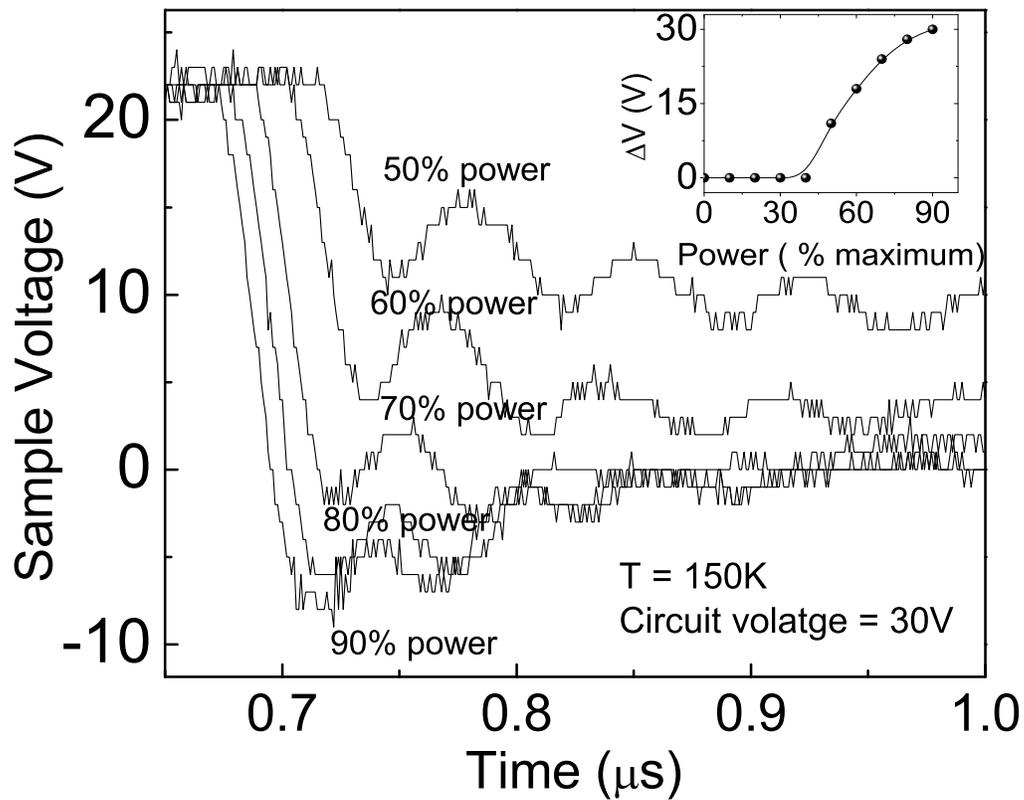}%\
\caption{\label{fig17}The variation of the sample voltage in real time for different intensity of the 6 ns laser pulse. The
sample held at 150 K with applied bias voltage less than the critical voltage for current switching.  The inset shows the
maximum drop in voltage ($\Delta$V) that takes place across the sample on photoexposure as a function of laser power.}
\end{figure}

 \clearpage
\begin{figure}
\includegraphics [width=16cm]{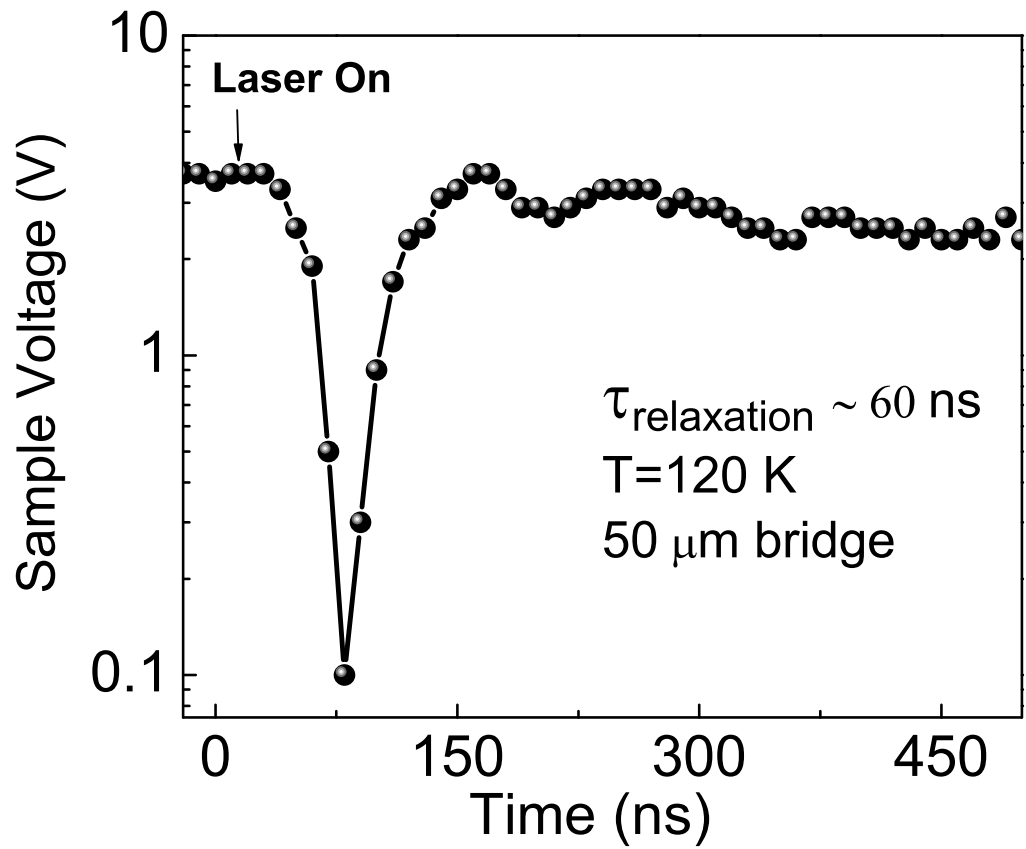}%\
\caption{\label{fig18}Real time variation of the sample voltage as a result of photoillumination at 120 K where the sample is
initially driven to the metallic state by applying an electric field. In this case the sample relaxes to its initial state in
about 60 ns. The solid lines are a guide to the eye. }
\end{figure}

\end{document}